**Title:** Heterogeneously integrated GaAs waveguides on insulator for efficient frequency conversion

*Lin Chang,[1,*] Andreas Boes,[1,2] Xiaowen Guo,[1] Daryl T. Spencer,[3] MJ. Kennedy,[1]
Jon D. Peters,[1] Nicolas Volet,[1] Jeff Chiles,[4] Abijith Kowligy,[3] Nima Nader,[4]
Daniel D. Hickstein,[3] Eric J. Stanton,[4] Scott A. Diddams,[3] Scott B. Papp,[3] John E. Bowers[1]*

*Corresponding Author: E-mail: linchang@umail.ucsb.edu

[1] Department of Electrical and Computer Engineering, University of California, Santa Barbara, CA 93106, USA
[2] School of Engineering, RMIT University, Melbourne, VIC 3000, Australia
[3] Time and Frequency Division, National Institute of Standards and Technology, Boulder, CO 80305 USA
[4] Applied Physics Division, National Institute of Standards and Technology, Boulder, CO 80305 USA

Tremendous scientific progress has been achieved through the development of nonlinear integrated photonics. Prominent examples are Kerr-frequency-comb generation in micro-resonators, and supercontinuum generation and frequency conversion in nonlinear photonic waveguides. High conversion efficiency is enabling for applications of nonlinear optics, including such broad directions as high-speed optical signal processing, metrology, and quantum communication and computation. In this work, we demonstrate a gallium-arsenide-on-insulator (GaAs) platform for nonlinear photonics. GaAs has among the highest second- and third-order nonlinear optical coefficients, and use of a silica cladding results in waveguides with a large refractive index contrast and low propagation loss for expanded design of nonlinear processes. By harnessing these properties and developing nanofabrication with GaAs, we report a record normalized second-harmonic efficiency of 13,000% $W^{-1}cm^{-2}$ at a fundamental wavelength of 2 µm. This work paves the way for high performance nonlinear photonic integrated circuits (PICs), which not only can transition advanced functionalities outside the lab through fundamentally reduced power consumption and footprint, but also enables future optical sources and detectors.

## 1. Introduction

Nonlinear optics has been an important branch of optics research since the groundbreaking demonstration of second-harmonic generation (SHG) in 1961 [1]. The origin of nonlinear-optical effects lies in material's nonlinear dielectric polarization, which responds to the incident optical field and can generate new optical carriers. The generation of new optical frequencies is powerful and has been widely used. It enabled a vast array of capabilities in optical signal generation and processing, such as switching and demultiplexing of signals at unprecedented speeds [2,3], ultrashort pulse measurement and generation [4,5], optical synthesizers, clocks, and radiofrequency spectroscopy at terahertz speeds [6-9]. In the field of quantum computation and communication, nonlinear optical components are widely used to generate entangled photon pairs [10] and to convert the frequency of single photons to telecommunication wavelengths [11]. This breath of applications motivates the creation of efficient nonlinear optical components with photonic integrated circuits (PICs), reducing the cost, footprint and power consumption [12-14].

A natural approach to improve the nonlinear optical device performance is to use materials with high optical nonlinear coefficients. GaAs and the closely related AlGaAs alloy are very promising materials [15-21], as they have one of the largest second ($\chi^{(2)}$) and third order nonlinear optical coefficients ($\chi^{(3)}$), orders of magnitude higher than those of other commonly used nonlinear optical materials (see Table 1). So far, most GaAs (AlGaAs) waveguides used GaAs (AlGaAs) thin films on native substrates due to the epitaxial growth requirements [17]. Those waveguides have a relative weak vertical refractive index contrast ($\Delta n \approx 0.2$), which limits the achievable optical intensity and hampers waveguide designs that fulfil the phase matching condition or allow dispersion engineering. Several techniques have been demonstrated to overcome these issues, such as thermally oxidizing of AlGaAs cladding layers [18, 19] or suspending the waveguides [20, 21]. However, those approaches suffer from high optical waveguide losses, which prevent the use of such designs for long waveguides or high Q resonators to boost the nonlinear process. Furthermore, most of these waveguide designs are difficult to integrate in common PICs platforms. These limitations can be overcome by heterogeneous integration, which has recently attracted a lot of interest as it enables a convenient way to integrate high quality nonlinear materials into PICs [14, 15]. Heterogeneous integration also enables the use of low-index claddings, e.g. $SiO_2$ for high optical confinement, which enhances the nonlinear optical interaction due to the higher intensity.

In this work, we present a GaAs waveguide nonlinear optical platform, which is fully cladded with $SiO_2$. The waveguide structure is achieved by heterogeneous bonding the GaAs onto an oxidized Si substrate and coating it with $SiO_2$. This platform enables highly efficient nonlinear optical processes, thanks to the high material nonlinearity of GaAs, the strong refractive index contrast and the low propagation loss of the waveguides. By using this approach, we achieved a record high SHG normalized efficiency of 13,000% $W^{-1}cm^{-2}$ at a fundamental wavelength of 2 µm, which is one order of magnitude higher than previously demonstrated nonlinear optical waveguide devices [13, 18].

**2. Device design**

Efficient SHG requires the fulfillment of the phase matching condition. This means that the phase relationship between the interacting waves (pump and SH light waves) are maintained along the propagation direction. Common methods to achieve phase matching in bulk crystals are birefringent and quasi-phase matching (QPM), both of which have been widely applied in the literature [22, 31]. For chip scale PICs, the waveguide dimensions provide an additional degree of freedom to achieve phase matching. This can be explained by the fact that for waveguides with submicron dimensions, the dispersion is mainly determined by waveguide geometry rather than the material dispersion. In this case the effective refractive indices for different polarized modes of the waveguide at the pump and SH wavelength can be matched by finding the right waveguide geometry (modal phase matching) [21].

The schematic cross section of a GaAs waveguide in our platform is shown in Fig. 1 (a). Here, we designed a waveguide to achieve SHG when using a fundamental pump wavelength of 2 µm. Fig. 1 (b) and (c) show the simulated mode distributions for the fundamental TE and TM modes at 2 and 1 µm wavelength, respectively. The effective refractive indices of the two modes is matched by tailoring the thickness and width of GaAs waveguide. Fig. 1 (d) shows the relation between the waveguide thickness and width to achieve phase matching for a pump wavelength of 2 µm – a thicker GaAs film requires a wider waveguide to fulfill the phase matching condition. For our purposes, we chose a GaAs thickness of 150 nm, which corresponds to a waveguide width of 1.5 µm. This is a compromise between a relatively narrow waveguide to achieve a high intensity, while not too small to cause increased propagation loss due to scattering from the waveguide sidewalls. Fig. 1 (e) shows the effective indices of the two modes as a function of waveguide width. Phase matching is fulfilled at the intersection point of the two curves.

The calculated normalized efficiency for the chosen waveguide geometry is 32,000% $W^{-1}cm^{-2}$. This value is one order of magnitude higher than previous reported numbers of thin film $LiNbO_3$ platform [13, 31]. This significant increase is mainly caused by two factors: the higher nonlinear coefficient and the smaller mode size. When one compares the GaAs platform to the thin film $LiNbO_3$ waveguide platform, it can be found that the GaAs waveguide has a 4-6 times higher $\chi^{(2)}$ ($d_{14}$) and 4 times smaller waveguide mode size. In addition, the modal phase matching gains a factor of $(\pi/2)^2$ [22] enhancement in efficiency when compared to QPM, because the SHG light is in phase with the pump wavelength along the whole waveguide length.

## 3. Fabrication

The device fabrication process is illustrated in Fig. 2. We used a GaAs chip diced from a wafer prepared by metal-organic chemical vapor deposition (MOCVD). The layer structure is shown in Fig. 2: a [001] orientated 150-nm-thick GaAs film and a 500-nm-thick $Al_{0.8}Ga_{0.2}As$ layer were grown on a 500-µm-thick GaAs substrate. A 5-nm-thick SiN layer was sputtered on the GaAs thin-film surface, to enhance the bonding strength compared to $SiO_2$-GaAs interface. This chip was bonded onto a Si wafer with 3-µm-thick thermal $SiO_2$ layer, after plasma activation [32]. The thermal $SiO_2$ layer was patterned before the bonding by inductively coupled plasma (ICP) etching to form 5 × 5 µm$^2$ square vertical channels (VCs) with 50 µm spacing for gas release. The bonded piece was annealed at 200°C for 12 hours under pressure to enhance the bonding strength. Afterwards, mechanical polishing was applied to lap the GaAs substrate thickness down to 70 µm. The remaining GaAs substrate was removed by wet etching with $H_2O_2$:$NH_4OH$ (30:1) and the

Al$_{0.8}$Ga$_{0.2}$As layer was removed by buffered hydrofluoric acid (BHF). Fig. 3 (a) shows the picture of the bonded chip after substrate removal, with a bonding yield larger than 95%. The surface roughness of the bonded GaAs thin film on the chip is ~0.3 nm (RMS), which is very similar compared to the pre-processing surface of GaAs (~0.2 nm), both measured by atomic force microscopy.

After substrate removal and AlGaAs wet etch, a layer of SiO$_2$ was deposited on the GaAs thin-film. Inductively Coupled Plasma (ICP) etching was used to pattern the SiO$_2$ layer, which acts as a hard mask for a second ICP etching step that patterns the GaAs layer. Fig. 3 (b) shows an SEM picture of the waveguide after GaAs etching, indicating smooth sidewalls. Finally, the sample was coated with another layer of SiO$_2$ to protect the waveguides. The final waveguide cross section is shown in Fig. 3 (c).

## 4. Experimental results and discussions

A schematic illustration of the nonlinear optical characterization setup is shown in Fig. 4 (a). A 1975-2075 nm tunable CW laser (New Focus TLB6700) is used as light source. About 1 mW (0 dBm) power from the laser's free-space output is coupled into a 2 µm wavelength single mode fiber. The fiber coupled light passes through a 2 µm fiber amplifier (AdValue Photonics), which increases the pump power. A polarization controller is used afterwards to align the polarization of the light so that the fundamental TE mode of the GaAs waveguide is excited when using a lensed fiber. The light from the waveguide output port is collected by another lensed fiber, which is connected to a wavelength-division multiplexer, splitting the pump and SHG light. The pump light is analyzed by an optical spectrum analyzer (OSA) and the SH power is measured by a Si photodetector, respectively.

The SHG characterization results for a 1.4-mm-long, 1.53-µm-wide waveguide are plotted in Fig. 4 (b). The input port of the waveguide device is a 200-µm-long, 350-nm-wide waveguide, connected to the SHG waveguide section by a 100-µm-long linear taper. This was done to reduce the coupling loss at input port, which is estimated to be around 0.45 (3.5 dB). The coupling losses for the pump and SHG light at the output port, which is a normal edge coupler with same waveguide width as the SHG section, are estimated to be around 0.75 (6 dB) and 0.7 (5.2 dB), respectively. No inverse taper was chosen on this side, as it is difficult to fabricate the desired waveguide width (150 nm) for the SH wavelength by our current lithography, which can be solved in future by using electron beam lithography. The fabricated waveguide has a propagation loss at the pump wavelength of approximately 1-2 dB/cm. This value was extracted by using the Fabry-Perot method and verified by the quality factor of a ring resonator with same waveguide geometry. The low waveguide loss is a result of the smooth waveguide sidewalls (see Fig. 3 (b)), top and bottom surface (RMS ~0.3 and 0.2nm). Fig. 4 (b) presents the pump ($P_\omega$) and SHG ($P_{2\omega}$) power as a function of the pump wavelength, where the plotted powers refer to the power levels inside of the waveguide. Fig. 4 (c) shows the normalized efficiency of this waveguide (Red dots), which is extracted based on the formula $P_{2\omega}/(P_\omega L)^2$. The maximum single pass conversion efficiency of this waveguide is about 250% $W^{-1}$, which corresponds to a normalized efficiency of 13,000% $W^{-1}cm^{-2}$ for a 1.4 mm long waveguide. The blue curve shows a plot of $sinc^2$ function fitted to experimental result, which indicates that the spectral shape of the measurement result closely matches the theoretically expected function. The fitted full width at half maximum (FWHM) bandwidth of the

sinc$^2$ function is 0.93 nm, which is very close to the theoretical prediction of 0.90 nm, indicating good agreement.

The transmission spectra of the fundamental wavelength in Fig. 4 (b) shows periodic ripples, caused by the waveguide end faces, which form a low-finesse Fabry-Perot cavity. This resonance enhances the fundamental power and therefore also the generated SH power inside of the waveguide as evident in Fig. 4 (b) by the periodic ripples in the fundamental and SH power. The extinction ratio of the Fabry-Perot ripples is ~1 dB, which corresponds to a 0.5 dB enhancement of peak power compared to the power coupled into waveguide. As a result, the resonance enhanced SHG efficiency (~300% W$^{-1}$) is about 1 dB higher than the single pass efficiency. By increasing the reflectivity of the facets, the resonance enhanced conversion efficiency can even further be improved due to the low propagation loss of the waveguide.

The normalized efficiency that we achieved (13,000% W$^{-1}$cm$^{-2}$) is about 2.5 times smaller compared to the simulated normalized efficiency (32,000% W$^{-1}$cm$^{-2}$) for the chosen waveguide geometry. Two possible factors may cause this discrepancy, beyond the experimental and theoretical uncertainties involved in calibrating these normalized efficiencies. One likely reason for this efficiency drop is the non-uniformity of the GaAs waveguide geometry, especially the variation of the GaAs thickness. Figure 5 (a) plots the phase matching wavelength of a 1.5-μm-wide waveguide as a function of different GaAs thicknesses. It can be seen that a 1 nm change in thickness causes an 11 nm shift in the phase matching wavelength. This indicates that the GaAs thickness must be extremely uniform over the propagation length in order to achieve high conversion efficiencies. The variation of waveguide width is another uniformity concern that can impact the nonlinear optical efficiency of the waveguide. According to Fig. 5 (b), a 10 nm change in width shifts the phase matching wavelength by 1 nm. Another possible reason is the high propagation loss of light at a wavelength of 1 µm. Previous reports that used AlGaAs waveguides for SHG [19,21] found a ten times higher propagation loss at the SH wavelength, compared to the loss at fundamental wavelength. In Fig. 5 (c) we plotted the dependence of efficiency for a 1.4-mm-long waveguide as a function of the propagation loss at a wavelength of 1 µm. It can be seen that the loss at the SH wavelength can cause a significant drop in the conversion efficiency for loss levels in the order of tens of dB/cm. It should be also noted that there is an uncertainty about the nonlinear coefficient of GaAs in literature. We used the Miller's rule to estimate the nonlinear optical coefficient at a wavelength of 2 µm from the measurements reported in Refs. [24,25], which seem to be fairly consistent with each other. However, older references report even higher nonlinear optical coefficients of GaAs [33].

The device conversion efficiency can be further improved by applying a ring resonator geometry to the waveguide structure. Using a ring resonator will enhance the intensity in the waveguide, boosting the nonlinear optical process at the expense of a narrow bandwidth. Furthermore, the ring resonator has smaller footprint compared to a straight waveguide, relieving the non-uniformity thickness issue of the GaAs thin-film. However, one needs to keep the crystal symmetry of GaAs in mind to achieve phase matching in resonators. This requires that the two refractive indices of the pump and SH modes are QPM rather than direct phase matched, similar with the work in Ref. [34]. A race-track resonator structure can also be used, where direct phase matching is utilized in the straight sections of the device.

Another way to improve the conversion efficiency is to apply High-Reflection (HR) coating on both end faces of a straight waveguide. We calculated that the resonance enhanced conversion efficiency can be increased to a staggering 1,000,000%W$^{-1}$, by adding 90% and 99% HR coating on the input and output faces of the 1.4-mm-long waveguide from Fig. 4(d), respectively. Coupling power in such a resonate structure can be difficult, which to a certain point may decrease the external efficiency. Overcoming this issue might require the integration of a light source inside the cavity, for example by heterogeneous integration [35, 36] or epitaxial growth [37]. Such a structure will be able to significantly convert a micro-Watt pump power to other frequencies. This will be very useful in building light source at various wavelengths, quantum-related applications, direct self-reference of frequency comb and many other on-chip applications.

Moreover, AlGaAs can also be used to build similar structures for SHG that can operate below a pump wavelength of 1700 nm, by shifting the bandgap of the material to shorter wavelengths and avoid material absorption. Thicker GaAs (AlGaAs) waveguides, which satisfy the anomalous dispersion requirement, are valuable in frequency comb and supercontinuum generations [15]. The propagation loss of thicker waveguides is expected to be lower than that of the current waveguides, which may enable ring resonators with a high quality factor (up to millions), reducing the threshold for comb generation to micro-Watt levels thanks to the high $\chi^{(3)}$.

## 5. Conclusion

In conclusion, we successfully demonstrated a heterogeneous GaAs nonlinear optical platform that is fully cladded with SiO$_2$ on a Si wafer. This platform provides the highest nonlinear coefficients of $\chi^{(2)}$ and $\chi^{(3)}$ among commonly used nonlinear optical platforms, a high refractive index contrast and a great flexibility for phase matching and dispersion engineering. We demonstrated SHG in this platform, with a record efficiency of 13,000% $W^{-1}cm^{-2}$. This platform paves the way for previously inaccessible nonlinear optical experiments due to its high efficiency. It also has a great potential to be heterogeneously integrated in PICs, due to the low pump power requirements, small footprint, and the capability to be fully integrated with active devices.

**Funding** DARPA MTO DODOS contract (HR0011-15-C-055).

**Acknowledgements** We thank Daehwan Jung, Justin Norman, Chen Shang, Tony Huang, Minh Tran, Tin Komljenovic, Paolo Pintus, Alfredo J. Torres and Alexander Spott for valuable discussions and help in fabrications. We would also like to thank Richard Mirin, Martin Fejer, Carsten Langrock and Alireza Marandi for the fruitful discussions. N.V. acknowledges support from the Swiss National Science Foundation.

**Keywords**: nonlinear optics, integrated photonics, silicon photonics, wavelength conversion devices.

**Figure 1.** Nonlinear waveguide design: (a) Waveguide cross section geometry; (b) Mode distribution of fundamental TE mode at 2 µm wavelength; (c) Mode distribution of fundamental TM mode at 1 µm wavelength; (d) Required waveguide width to achieve phase matching for SHG at a fundamental wavelength of 2 µm as a function of the GaAs thickness; (e) Effective indices of the pump and SH modes as a function of the waveguide width for a GaAs thickness of 150 nm at wavelengths of 2 µm and 1 µm, respectively.

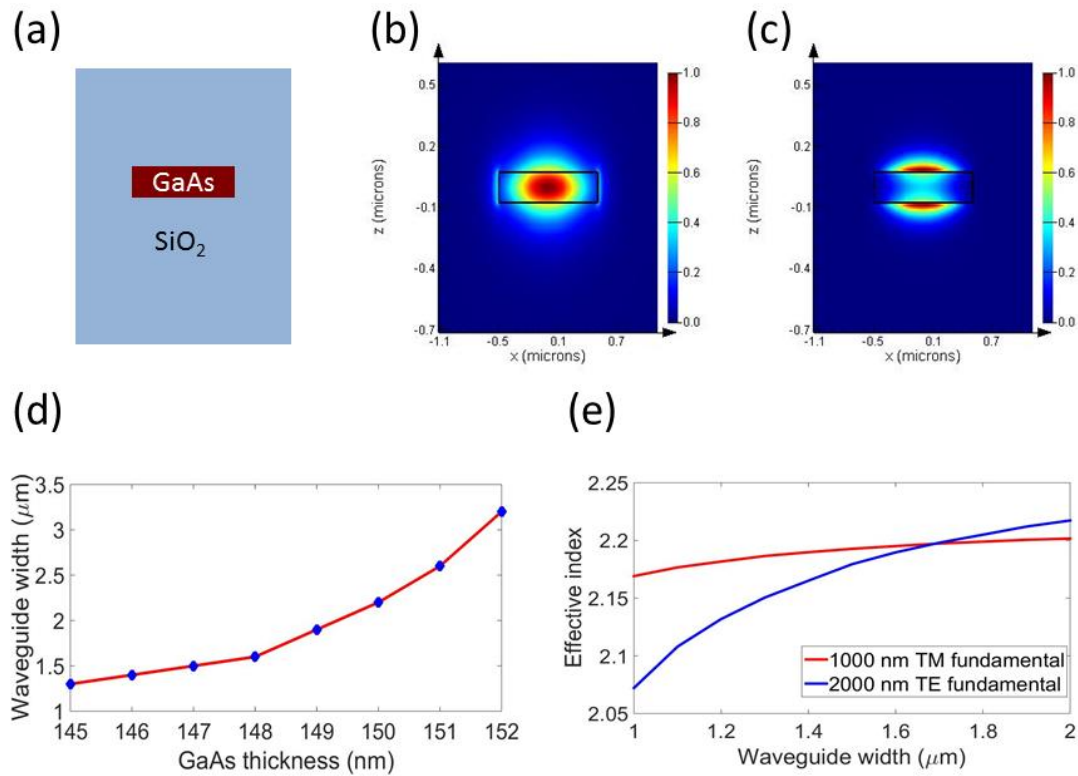

**Figure 2.** Process flow for GaAs waveguide device fabrication.

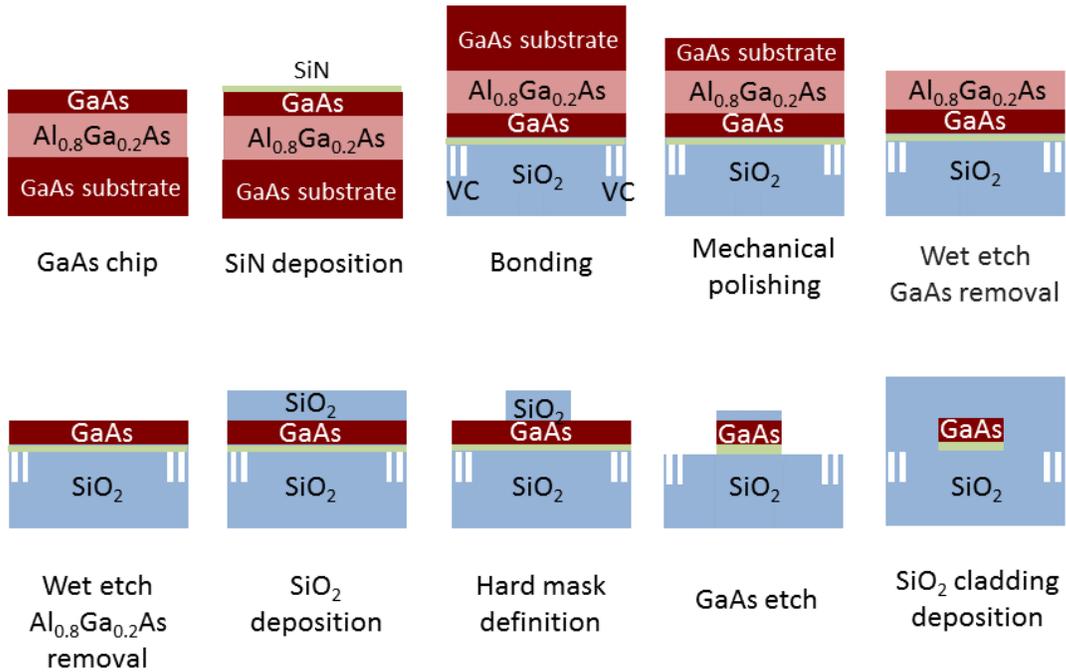

**Figure 3.** Device images: (a) Bonded GaAs thin film on $SiO_2$ after substrate removal; (b) SEM image of GaAs waveguide with $SiO_2$ hard mask on top after GaAs etch; (c) SEM image of the waveguide cross section.

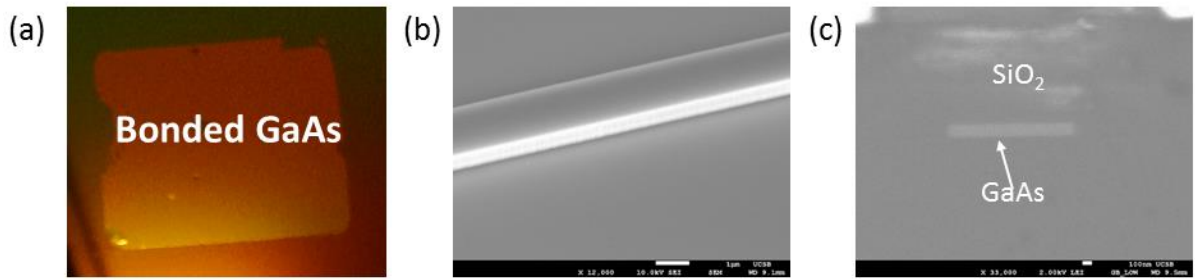

**Figure 4.** (a) Schematic of the SH characterization setup; (b) SHG and pump power as a function of the pump wavelength for a 1.4-mm-long waveguide with inverse taper input; (c) Single pass conversion efficiency extracted from (b).

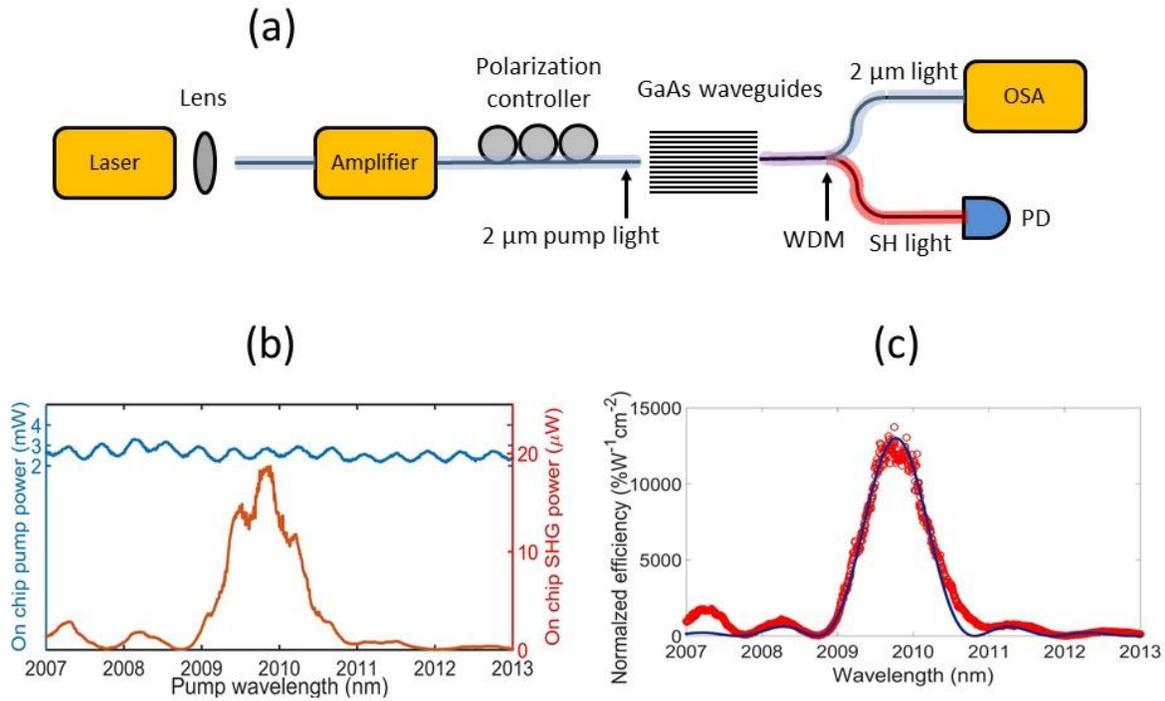

**Figure 5.** Dependence of phase matching wavelength on the GaAs waveguide (a) thickness and (b) width; (c) Calculated efficiency over propagation loss of SH light for 1.4-mm-long waveguide, assuming 2 dB/cm loss at pump wavelength.

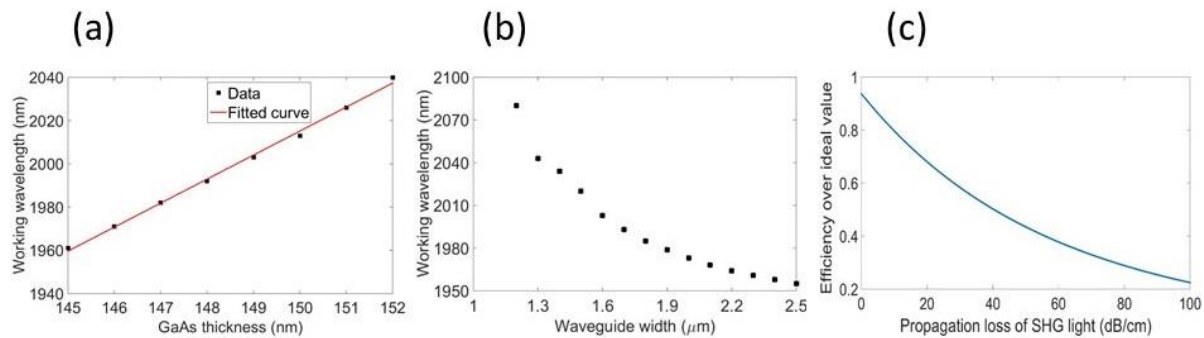

**Table 1.** Comparison of nonlinear optical coefficients and mode sizes of waveguides among commonly used nonlinear materials.

| Material | $d$ ($\frac{1}{2}\cdot\chi^{(2)}$) [pm/V] | $\chi^{(3)}$ [cm$^2$/W] | Mode size [μm$^2$] |
|---|---|---|---|
| LiNbO$_3$ | 30 [22] | $5.3 \times 10^{-15}$ [26] | ~2 |
| AlN | 1 [23] | $2.3 \times 10^{-15}$ [27] | ~1 |
| Si$_3$N$_4$ | - | $2.5 \times 10^{-15}$ [28] | ~1 |
| Si | - | $6.5 \times 10^{-14}$ [29] | ~0.5 |
| GaAs (AlGaAs) | 119 [24,25] | $1.6 \times 10^{-13}$ [30] | ~0.5 |